\documentstyle[12pt]{article}
\input epsf

\begin{document}

\begin{titlepage}

\begin{flushright}
UM--TH--98--02\\
LAPTH--674--98\\
Freiburg--THEP--98--02\\
February 1998
\end{flushright}
\vspace{2.5cm}

\begin{center}
\large\bf
{\LARGE\bf Higgs mass saturation effect and the LHC discovery potential}\\[2cm]
\rm
{A. Ghinculov$^{a,}$\footnote{Work supported by the US Department of Energy (DOE)}, 
 T. Binoth$^b$, J.J. van der Bij$^c$}\\[.5cm]

{\em $^a$Randall Laboratory of Physics, University of Michigan,}\\
      {\em Ann Arbor, Michigan 48109--1120, USA}\\[.2cm]
{\em $^b$Laboratoire d'Annecy--Le--Vieux de Physique 
         Th\'eorique\footnote{URA 1436 associ\'ee \`a l'Universit\'e de Savoie} LAPP,}\\
      {\em Chemin de Bellevue, B.P. 110, F-74941, 
           Annecy--le--Vieux, France}\\[.2cm]
{\em $^c$Albert--Ludwigs--Universit\"{a}t Freiburg,
                               Fakult\"{a}t f\"{u}r Physik,}\\
      {\em Hermann--Herder Str.3, D-79104 Freiburg, Germany}\\[3.cm]
      
\end{center}
\normalsize

\begin{abstract}
In view of recent perturbative and nonperturbative evidence 
that the peak of the Higgs resonance saturates as the coupling increases,
we examine the potential of the future Large Hadron Collider to 
discover a heavy Higgs resonance by gluon fusion.
\end{abstract}


\end{titlepage}


\title{Higgs mass saturation effect and the LHC discovery potential}

\author{A. Ghinculov$^{a,}$\thanks{Work supported by the US 
        Department of Energy (DOE)}, T. Binoth$^b$, J.J. van der Bij$^c$}

\date{{\em $^a$Randall Laboratory of Physics, University of Michigan,}\\
      {\em Ann Arbor, Michigan 48109--1120, USA}\\[.2cm]
      {\em $^b$Laboratoire d'Annecy--Le--Vieux de Physique 
         Th\'eorique\thanks{URA 1436 associ\'ee \`a l'Universit\'e de Savoie} LAPP,}\\
      {\em Chemin de Bellevue, B.P. 110, F-74941, 
           Annecy--le--Vieux, France}\\[.2cm]
      {\em $^c$Albert--Ludwigs--Universit\"{a}t Freiburg, 
               Fakult\"{a}t f\"{u}r Physik,}\\
      {\em Hermann--Herder Str.3, D-79104 Freiburg, Germany}}

\maketitle

\begin{abstract}
In view of recent perturbative and nonperturbative evidence 
that the peak of the Higgs resonance saturates as the coupling increases,
we examine the potential of the future Large Hadron Collider to 
discover a heavy Higgs resonance by gluon fusion.
\end{abstract}



No doubt, the main goal of the Large Hadron Collider is a better
understanding of the electroweak spontaneous symmetry breaking mechanism.
At present, the constraints on the mass of a standard Higgs boson obtained
from radiative corrections are quite loose, and depend strongly on which
data are taken into account \cite{degrassi}. Even values of the order 
of 1 TeV cannot
be excluded with certainty. For this reason one should keep an eye
open to the possibility of a heavy Higgs particle.
On the other hand, the LHC ought to be able to observe a standard Higgs
resonance in the mass range under $\sim$ 1 TeV \cite{aachenwork}. 

The physics of light
Higgs particles is well understood, and a lot of work was done on
radiative corrections. 
For a mass approaching the 1 TeV scale, the
situation is more complicated because the Higgs field becomes 
strongly coupled and the radiative corrections blow up.

Recent nonperturbative results \cite{oneovn} 
which involve a higher order $1/N$ expansion
of the $O(N)$--symmetric sigma model indicate an interesting behaviour
of the Higgs sector at strong coupling. As the coupling increases, a
saturation effect takes place. The mass of the Higgs resonance does
not increase beyond a saturation value under 1 TeV, and only its width
increases. 

\begin{figure}
\hspace{1.5cm}
    \epsfxsize = 13cm
    \epsffile{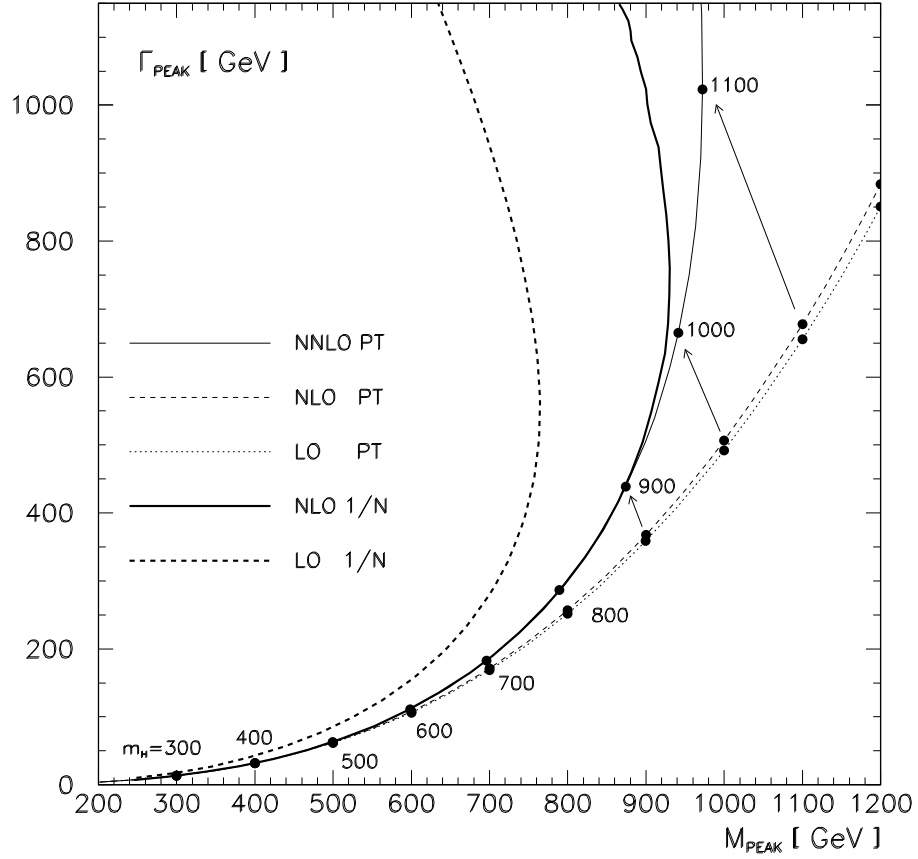}
\caption{{\em The saturation effect. The parameters $M_{PEAK}$
and $\Gamma_{PEAK}$ are extracted from the position and the height 
of the Higgs resonance in fermion scattering as if the resonance was 
of Breit--Wigner type. We give the relation between $M_{PEAK}$
and $\Gamma_{PEAK}$ in perturbation theory (LO, NLO and NNLO) 
and in the nonperturbative $1/N$ expansion (LO and NLO). For
the perturbation theory curves we give the corresponding values
of the on--shell mass parameter $m_H$.}}
\end{figure}

We illustrate the saturation effect in fig. 1. 
One can see that the saturation effect is suggested already by
the two--loop perturbative result, and is confirmed by the $1/N$ expansion.
In this figure we
plot an effective mass and width of the standard Higgs particle,
$M_{PEAK}$ and $\Gamma_{PEAK}$. These parameters  
are defined based on the fermion scattering process 
$f \bar{f} \rightarrow H \rightarrow f^{\prime} \bar{f^{\prime}}$.
$M_{PEAK}$ and $\Gamma_{PEAK}$ are extracted from the line shape 
of the Higgs resonance in this scattering process as if the resonance
was of Breit--Wigner type \cite{oneovn}. 

The perturbative results shown in fig. 1 are accurate in the
low coupling limit, and become progressively unreliable
numerically as the coupling increases. On the contrary, 
the convergence of the $1/N$ series is independent of the value of
the coupling; it only depends on the value of $N$, which is four
in the case of the standard model.
The striking feature of fig. 1 is that both expansions appear to 
converge towards a common result. In fact, the agreement between
the next--to--leading order $1/N$ result and two--loop perturbation
theory is remarkable. This convinces us that one has reached a good
numerical understanding of the behaviour of the Higgs sector
rather deep in the saturation zone, up to about $m_H \sim$ 1.1 TeV.

Of course, the mass and the width defined from the line shape 
of the resonance are process dependent. This is in contrast to the 
position of the pole in the complex plane
\cite{higgspole}, which is universal. 
Nevertheless, fig. 1 is conclusive for the occurrence of the
mass saturation effect, and for the degree of accuracy provided
by perturbation theory and the $1/N$ expansion in different
orders which are available.

It is clear from fig. 1 that
the properties of the Higgs particle differ considerably from the
perturbative leading order and one--loop results. The low 
order results can be substantially misleading 
in the range above 700---800 GeV. 

In view of the mass saturation
effect, it is the purpose of this
letter to reevaluate the gluon fusion process, which is the main  
Higgs production mechanism at LHC, and to estimate the LHC potential
to discover a strongly interacting Higgs resonance.
On the one hand, the mass saturation effect shifts the resonance
towards lower energy, which tends to increase the production rate.
On the other hand, also the width is increased, which decreases
the production rate.

The Higgs boson production by gluon fusion
is dominated by the top quark loop contribution.  
This process was studied in detail at leading order 
\cite{aachenwork,ggleading}.
The full $Z$ pair production process $pp \rightarrow ZZ$ is available, including
the coherent $gg \rightarrow ZZ$ and incoherent $q\bar{q} \rightarrow ZZ$
backgrounds to Higgs production. The incoherent background is meanwhile known to
order ${\cal O}(\alpha_s)$ \cite{mele.et.al}.
The perturbative corrections of enhanced electroweak 
strength up to NNLO were calculated in ref. 
\cite{gg2loop}. The next--to--leading
order nonperturbative $1/N$ result for the resonant diagram 
$gg \rightarrow H \rightarrow ZZ$ is in principle already contained
in the results of refs. \cite{oneovn}\footnote{Notice that 
the next--to--leading
order $1/N$ correction to the $H \rightarrow zz$ decay width is given by
the imaginary part of the {\em next--to--leading} order Higgs self--energy.
This is in contrast to perturbation theory, where the imaginary part of
the self--energy needs to be calculated one loop higher 
than the vertex correction.}.
Nevertheless, one notes that this is not sufficient for calculating the
process $gg \rightarrow ZZ$ nonperturbatively in the Higgs coupling
because also the phase factor of the decay amplitude $H \rightarrow ZZ$
is needed. This is because of interference effects with the 
nonresonant box diagrams. To perform this calculation one needs therefore
the next--to--leading order $1/N$ correction to the $Hzz$ vertex, which
is not available yet. Therefore, for the purposes of this letter, we
rely on the NNLO perturbative result in the on--shell 
renormalization scheme. We expect this to be a sensible approximation
up to a coupling corresponding to $m_H \sim$ 1.1 TeV
due to the good agreement with the NLO $1/N$ width (fig. 1).

In the following we consider the gluon fusion process
$gg \rightarrow H \rightarrow ZZ$, with the subsequent $Z$ 
pair decay into purely leptonic channels, which provide 
a clean signal. We will not deal with jet channels in this
paper; they have a larger branching ratio and thus have the 
potential of probing heavier Higgs resonances, but the 
analysis is complicated by the presence of a heavy QCD 
background and is strongly dependent on the detector's energy
and position resolution.
Therefore we will consider for the decay of the $Z$ pair
only the channels $(l^+l^-) (l^+l^-)$ and 
$(l^+l^-) (\nu \bar{\nu})$, where $l$ can be either the electron 
or the muon. We note that compared to the four muon channel, 
$(l^+l^-) (l^+l^-)$ has a branching ratio 4 times larger,
and $(l^+l^-) (\nu \bar{\nu})$ 24 times larger.

In the $(l^+l^-) (l^+l^-)$ channel the Higgs mass can be reconstructed
completely, and the Higgs boson is observed as a peak in the
invariant mass of the $Z$ pair. In the  $(l^+l^-) (\nu \bar{\nu})$
channel the Higgs particle is observed as a Jacobian peak in the 
missing transverse momentum distribution. Details of the 
leading order gluon fusion calculation can be found in ref. 
\cite{ggleading}. The NNLO
radiative corrections of enhanced electroweak strength can be 
derived from existing two--loop calculations \cite{calc2loop}
by using the procedure explained in ref. \cite{gg2loop}.
We work in the on--shell renormalization scheme to 
facilitate comparison with existing calculations.
The radiative corrections were 
incorporated into a Monte Carlo program which calculates the
complete $ZZ$ production in $pp$ collisions with the subsequent
$Z$ decay taken into account in the narrow width approximation. 

We plot in fig. 2 the distributions for the invariant mass $m_{ZZ}$
of the $Z$ pair and for the transverse momentum $p_t$ of the $Z$ 
bosons which can be constructed from the purely leptonic channels.
We considered the CM energy of the LHC $\sqrt{s}= 14$ TeV. We 
used the parton distribution functions of Martin, Roberts and Stirling 
\cite{mrs}. For a crude simulation of the detector acceptance,
we impose a rapidity cut on the final state charged leptons 
$\eta < 2.5$, and also require a minimum transverse momentum
for the charged leptons of 20 GeV.

\begin{figure}
\hspace{1.5cm}
    \epsfxsize = 13cm
    \epsffile{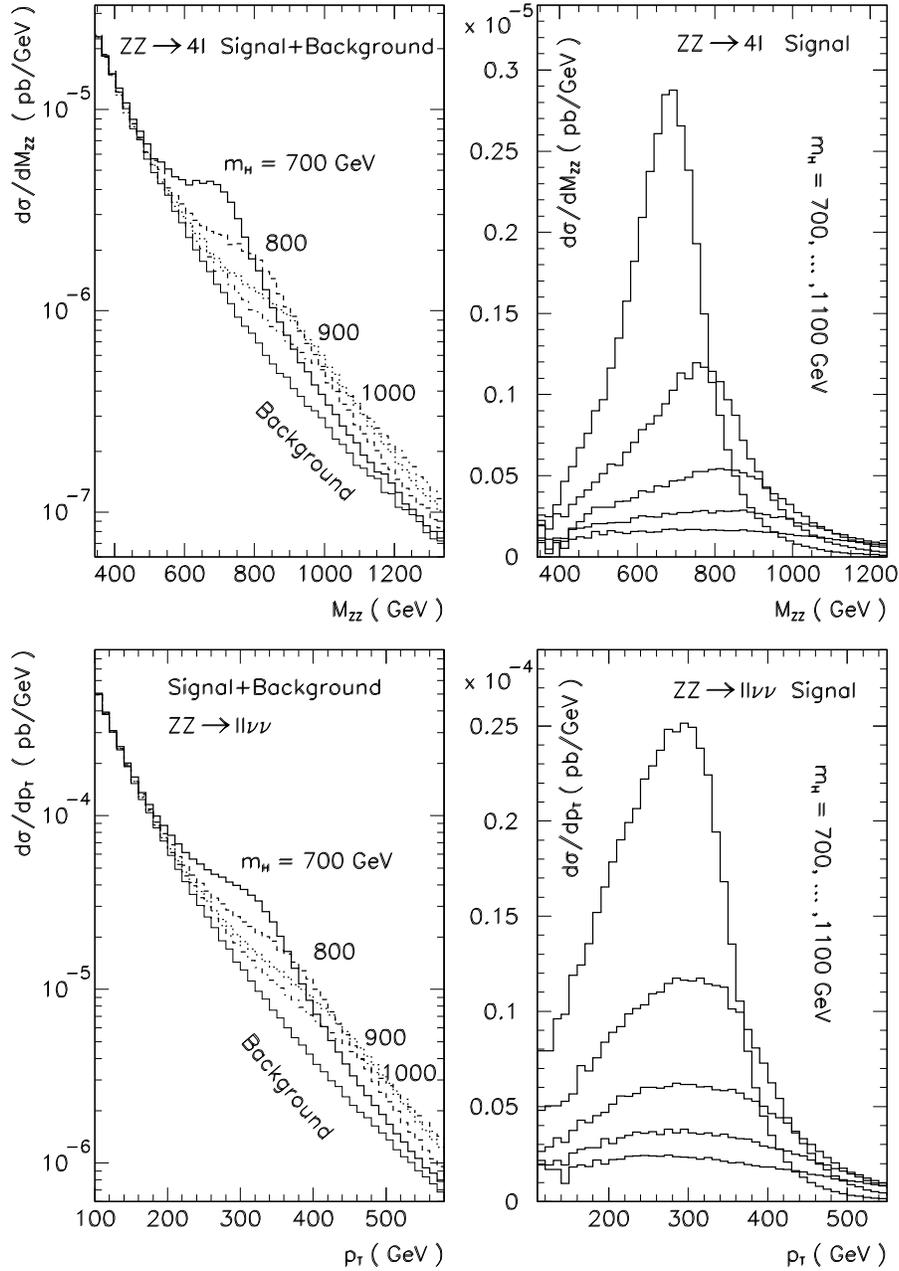}
\caption{{\em Distributions of the invariant mass of the $Z$
pairs (above) and of the transverse momentum of the $Z$ bosons (below).
We show both the full cross section and the Higgs signal after the
subtraction of the background. The background is defined by the absence
of the resonant Higgs production diagram.
For the $m_{ZZ}$ distribution we use the branching ratio 
of the $(l^+l^-) (l^+l^-)$ channels,
and for the $p_t$ distribution we use the 
$(l^+l^-) (\nu \bar{\nu})$ channel.
The LHC CM energy is 14 TeV, and $m_t=$ 180 GeV.}}
\end{figure}

Figure 2 shows the effect of the mass saturation on the
gluon fusion process. As the coupling is increased, the position 
of the peak of the distributions barely changes, but the resonance 
becomes wider. 

As the Higgs coupling is increased, the signal becomes weaker,
and the resonance becomes more difficult to observe.
To fully exploit the discovery potential of the machine, it
is of advantage to isolate the whole resonance region by cuts
and to compare the event rate with the expected background.
In the following we will assume an integrated luminosity of 
100 fb$^{-1}$. We impose suitable cuts separately on the
invariant mass of the $Z$ pair and on the transverse momentum
of the $Z$ bosons, and give the background and the signal for
different values of the on--shell mass parameter $m_H$ in table 1.
For comparison, we give also the event rates corresponding to the
measurement of the Jacobian peak of the transverse momentum of the
$Z$ bosons as it can be observed in the $(l^+l^-) (l^+l^-)$ 
events, but note that this is not competitive with the 
invariant mass analysis of this decay channel.

\begin{table}
\begin{center}
\begin{tabular}{||c||c|c|c|c|c|c|c||}           \hline\hline
 $m_{H}$ $[$GeV$]$               & 700  &  800  &  900 & 1000 & 1100 & 1200 & background  \\ \hline\hline
 $4l$: $m_{ZZ}$ distrib.         & 103  &  83   & 70   & 62   & 57   & 54   & 47          \\ 
 (additional $p_t$ cut)          & (87) &  (68) & (56) & (49) & (44) & (41) & (34)        \\ \hline
 $4l$: $p_t$ distrib.            & 81   &  64   & 53   & 46   & 41   & 38   & 32          \\ \hline\hline
 $2l2\nu$: $p_t$ distrib.        & 546  & 443  & 373  & 328  & 298  & 279  & 240          \\ \hline\hline
\end{tabular}
\end{center}
\caption{The number of events to be expected from a 100 $fb^{-1}$ sample by imposing
a cut 582 GeV $<m_{ZZ}<$ 1372 GeV on the $m_{ZZ}$ distribution of $(l^+l^-) (l^+l^-)$
events; 240 GeV $<p_t<$ 590 GeV on the $p_t$ distribution of $(l^+l^-) (l^+l^-)$;
and 240 GeV $<p_t<$ 590 GeV on the $p_t$ distribution of
$(l^+l^-) (\nu \bar{\nu})$ events. 
In brackets we give the event rates corresponding to $(l^+l^-) (l^+l^-)$ events
when the cut $p_t > m_{ZZ}/4 {\protect \sqrt{1-4 m^2_{Z}/m^2_{ZZ}}}$ is imposed in
addition to 582 GeV $<m_{ZZ}<$ 1372 GeV.
For the outgoing charged 
leptons we request $\eta<$ 2.5, and a transverse momentum larger than 20 GeV. 
The LHC CM energy is 14 TeV, and $m_t=$ 180 GeV.}
\end{table}

\begin{figure}
\hspace{1.5cm}
    \epsfxsize = 13cm
    \epsffile{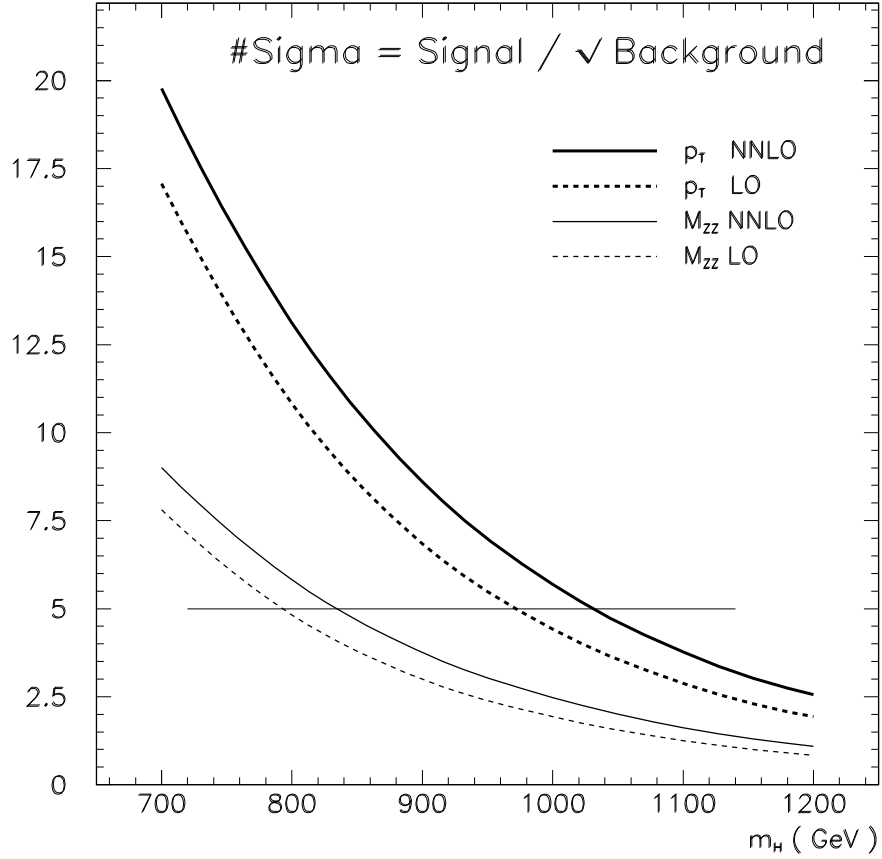}
\caption{{\em Discovery limits of a heavy Higgs resonance at LHC.
We give the limits which can be obtained from the invariant mass of the $Z$
pairs ($(l^+l^-) (l^+l^-)$ channel), 
and from the transverse momentum of
the $Z$ ($(l^+l^-) (\nu \bar{\nu})$ channel).
We use the cuts 582 GeV $<m_{ZZ}<$ 1372 GeV and 
240 GeV $<p_t<$ 590 GeV, respectively, to isolate the resonance
from each distribution. 
For the $(l^+l^-) (l^+l^-)$ channel we include the cut of eq. 1.
We also show the 
leading order results (dashed lines).
The LHC CM energy is 14 TeV, and $m_t=$ 180 GeV.}}
\end{figure}

For the $(l^+l^-) (l^+l^-)$ channel it has been proposed \cite{ggleading}
to use an additional cut:

\begin{equation}
  p_t > \frac{m_{ZZ}}{4} \sqrt{1 - 4 \frac{m_Z^2}{m_{ZZ}^2}}
\end{equation}
in order to improve the signal to background ratio of the
$m_{ZZ}$ distribution. This idea is based on the observation that
the Higgs boson decays isotropically in the center of mass frame,
while the background $Z$ pairs are radiated mainly at low transverse momentum.
We give in table 1, in brackets, the results obtained by using this cut. 
It can be seen that this cut improves to some extent the 
signal to background ratio. 

Keeping in mind that we have taken into account the
properties of the detector only by a rapidity cut as
explained above, the discovery potential of the LHC is summarized in fig. 3.
It can be seen that by observation of the neutrinoless leptonic
channels, and for a five sigma effect,
one can reach the zone up to about $m_H \sim$ 830 GeV,
where the saturation effect only starts to play a role.
As expected, the discovery potential is increased by 
using the missing $p_t$ channels, and the mass
saturation zone becomes accessible up to about $m_H \sim$ 1030 GeV
for a five sigma effect. Roughly speaking, the inclusion of quantum
effects increases the discovery range by 40--60 GeV with these cuts.
It can be expected that by using jet decay modes the discovery
range can be increased further in the saturation region, but 
then a careful analysis of the QCD background is needed. 

At this point we would like to make a few remarks regarding
the interplay of electroweak and QCD corrections. 
The one--gluon corrections to the Higgs boson production
by gluon fusion are available 
\cite{spira} for a narrow Higgs resonance. It has been shown 
that these corrections can be quite large, at the 
50\% --- 70\% level. This final enhancement effect
is the result of several contributions from virtual and real
gluon diagrams, some of them tending to enhance the cross section,
and other to decrease it. It is hard to infer from these
results even the sign of the effect for a wide resonance
because the full $gg \rightarrow ZZ$ process must be 
considered. This includes the one--gluon corrections
to nonresonant quark box diagrams, which are not available 
so far. Given the size of the
narrow width QCD corrections, it can be expected
that they be substantial for wide resonances as well.

To conclude, we reached a good quantitative understanding 
of electroweak effects for
heavy Higgs resonances over the whole range relevant for the LHC,
by combining two--loop perturbation theory and
next--to--leading order nonperturbative $1/N$ expansion.
The mass saturation effect is important for 
the gluon fusion Higgs signal at LHC for values of the on--shell
mass above 700---800 GeV. By including the leptonic missing $p_t$
decay modes one can probe into the mass saturation region.
Given that the identification of a heavy Higgs resonance depends
crucially on a good theoretical calculation of the background,
better control is needed of the interplay 
of electroweak and QCD corrections for a wide resonance.


\vspace{.5cm}

{\bf Acknowledgements}

One of us (A.G.) would like to acknowledge useful discussions
with Timo van Ritbergen. T.B. would like to thank J.--P. Guillet,
D. Buskulic and P. Aurenche for discussions.


\newpage


\end{document}